\documentclass[12pt]{article}  
\usepackage{bm}    
\usepackage{graphicx}  
\usepackage{amsmath}  
\usepackage{bm}
\usepackage{subfigure}
\usepackage{mathrsfs}

\begin{document}    




\begin{center}
\LARGE\bf Improvements in continuum modeling for biomolecular systems$^{*}$   
\end{center}

\footnotetext{\hspace*{-.45cm}\footnotesize $^*$Project supported by the China NSF (91230106) and the CAS Program for Cross \& Cooperative Team of the Science \& Technology Innovation.}
\footnotetext{\hspace*{-.45cm}\footnotesize $^\dag$Corresponding author. E-mail: bzlu@lsec.cc.ac.cn}

\begin{center}
\rm Qiao Yu \ and  \ Lu Ben-Zhuo$^{\dagger}$
\end{center}

\begin{center}
\begin{footnotesize} \sl
State Key Laboratory of Scientific and Engineering Computing, Academy of Mathematics and Systems Science, National Center for Mathematics and Interdisciplinary Sciences, Chinese Academy of Sciences, Beijing 100190, China.   
\end{footnotesize}
\end{center}


\vspace*{2mm}

\begin{center}
\begin{minipage}{15.5cm}
\parindent 20pt\footnotesize
Modeling of biomolecular systems plays an essential role in understanding biological processes, such as ionic flow across channels, protein modification or interaction, and cell signaling. The continuum model described by the Poisson-Boltzmann (PB)/Poisson-Nernst-Planck (PNP) equations has made great contributions towards simulation of these processes. However, the model has shortcomings in its commonly used form and cannot capture (or cannot accurately capture) some important physical properties of biological systems. Considerable efforts have been made to improve the continuum model to account for discrete particle interactions and to make progress in numerical methods to provide accurate and efficient simulation. This review will summarize recent main improvements in continuum modeling for biomolecular systems, with focus on the size-modified models, the coupling of the classical density functional theory and PNP equations, the coupling of polar and nonpolar interactions, and numerical progress.
\end{minipage}
\end{center}

\begin{center}
\begin{minipage}{15.5cm}
\begin{minipage}[t]{2.3cm}{\bf Keywords:}\end{minipage}
\begin{minipage}[t]{13.1cm}
Poisson-Boltzmann equation, Poisson-Nernst-Planck equations, ionic size effects, density functional theory, coupling of polar and nonpolar interactions, numerical methods
\end{minipage}\par\vglue8pt
{\bf PACS: }
87.15.A-
\end{minipage}
\end{center}

\section{Introduction}  
Modeling of biomolecular systems containing ionic particles and biomolecules is an important approach in life science for investigating electrostatic properties, such as ion distributions, electrostatic free energy, reactive rate and ion current$^{[1-9]}$. Based on a mean field framework, the continuum model treats ions in solution as continuum distributions rather than discrete particles, avoiding large degrees of freedom in computation. In the general case, the Poisson-Nernst-Planck (PNP) equations are utilized for continuum modeling of biomolecular systems. In particular, in the equilibrium state, ion density is assumed to obey Boltzmann distribution, and the Poisson-Boltzmann (PB) model can describe electrostatic interactions in a solvated system. It is notable that the steady-state PNP equations can be reduced to the PB equation at equilibrium conditions$^{[10]}$.

Although the continuum model has wide applications and has achieved great success in predicting many thermodynamic properties, the facts that ions are regarded as point charges and the model does not take into account ionic volume exclusion and ion-ion correlations make it unfeasible for systems where these effects are pronounced$^{[11-24]}$. The continuum model can lead to a high unphysical concentration of counterions in the vicinity of the biomolecule and miss the phenomena of ionic layering, overcharging or charge inversion and like-charge attraction$^{[10, 18, 25, 26]}$.

In the past few decades, a large number of methods were developed to improve the continuum model in order to precisely simulate biomolecular systems and obtain more reasonable computational results. Ionic size effects are incorporated in the continuum model by taking into account volume exclusion, resulting in a size-modified model$^{[11-14, 16-18]}$. Coupling of the classical density functional theory (DFT) and PNP equations has also been proposed to improve the continuum model through inclusion of excess chemical potential, which is variation of excess Helmholtz free energy with respect to ion concentrations$^{[20-24, 27]}$. Biomolecular exclusion expressed by a characteristic function, as well as other polar and nonpolar interactions, are added to the free energy to obtain optimal volume and surface and minimum free energy$^{[28-30]}$. Apart from these modifications, some other improvements to account for effects neglected by the traditional continuum models will be briefly discussed.

Along with relevant developments in physics, computer science, and mathematics, numerical methods have also been developed to obtain more accurate and efficient results$^{[1, 31]}$. A new treatment on the boundary is provided to give high solution accuracy and parallelizations by MPI, GPU and the Cilk Plus system are used to increase the speed of computations.

The purpose of this review is to provide an overview of the continuum models, their recent improvements, and numerical progress for modeling the biomolecular community. In section 2, we discuss the continuum model governed by PB/PNP equations. The improvements and numerical progress are presented in section 3 and section 4, respectively. Finally, we provide the conclusions in section 5.

\section{Continuum model}  
\begin{figure}[htbp]
\centerline{\includegraphics[scale=0.5]{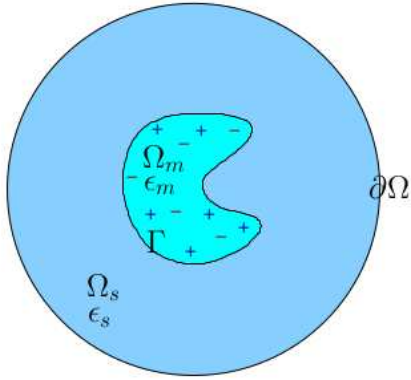}}
\caption{Illustration of a solvated biomolecular system.}
\label{solvated}
\end{figure}

In the simulation of solvated biomolecular systems, the computational region $\Omega$ is composed of two parts, the biomolecular region $\Omega_{m}$ with dielectric constant $\epsilon_{m}$, and the solution region $\Omega_{s}$ with dielectric constant $\epsilon_{s}$. Fig. \ref{solvated} schematically illustrates such a system, in which $\Gamma$ denotes the interface between the two different regions and $\partial\Omega$ stands for the outer boundary of the whole system$^{[1]}$.

The PB equation in $\Omega$ takes the form:
\begin{eqnarray}
    -\nabla\cdot\epsilon(\bm{x})\nabla\phi(\bm{x})-\sum_{i=1}^{K}\lambda c_{bi}e^{-\beta q_{i}\phi(\bm{x})}q_{i}=\rho^{f}(\bm{x}),\quad \text{in $\Omega$}, \label{PB}
\end{eqnarray}
where $\phi(\bm{x})$ is the potential needed to be solved, $\epsilon(\bm{x})=\epsilon_{\bm{x}}\epsilon_{0}$ represents the dielectric permittivity in which $\epsilon_{\bm{x}}$ is the relative permittivity at the given point $\bm{x}$, and $\epsilon_{0}$ is the vacuum permittivity, $K$ is the number of ion species in solution, the characteristic function $\lambda$ is 0 in $\Omega_{m}$ and 1 in $\Omega_{s}$, $c_{bi}$ is the bulk concentration of $i$th ion species with charge $q_{i}$, $\beta$ is defined by $\frac{1}{k_{B}T}$ where $k_{B}$ is the Boltzmann constant and $T$ is the absolute temperature, $\rho^{f}(\bm{x})=\sum_{i=1}^{N}Q_{i}\delta(\bm{x}-\bm{x_{i}})$, and $Q_{i}$ is the singular charge located at $\bm{x_{i}}$ representing one of the $N$ atoms from the biomolecule. On the interface $\Gamma$, two conditions need to be satisfied: (1) $\phi_{s}=\phi_{m}$ and (2) $\epsilon_{s}\frac{\partial\phi_{s}}{\partial n}=\epsilon_{m}\frac{\partial\phi_{m}}{\partial n}$, where $n$ is the outer unit vector of $\Gamma$. For the outer boundary $\partial\Omega$ far away from the biomolecule, the Dirichlet boundary condition is commonly used as $\phi|_{\partial\Omega}=0$.

In the general case, the PNP equations are given as follows:
\begin{eqnarray}
    \frac{\partial c_{i}(\bm{x})}{\partial t}=-\nabla\cdot J_{i}(\bm{x}),\quad \text{in $\Omega_{s}$},\quad i=1,\ldots,K,\label{NP}\\
    -\nabla\cdot\epsilon(\bm{x})\nabla\phi(\bm{x})-\rho^{ion}(\bm{x})=\rho^{f}(\bm{x}),\quad \text{in $\Omega$}, \label{P}
\end{eqnarray}
where
\begin{eqnarray}
    J_{i}(\bm{x})=-D_{i}(\bm{x})\Big(\nabla c_{i}(\bm{x})+\beta c_{i}(\bm{x})q_{i}\nabla\phi(\bm{x})\Big),
\end{eqnarray}
and $D_{i}(\bm{x})$ is the diffusive coefficient, $\rho^{ion}(\bm{x})=\sum_{i=1}^{K}c_{i}(\bm{x})q_{i}$, $c_{i}(\bm{x})$ is the concentration of the $i$th ion species at position $\bm{x}$. It's not difficult to find that the Boltzmann distribution can be derived from the Nernst-Planck equation (Eq. (\ref{NP})) at equilibrium conditions where $J_{i}(\bm{x})$ is zero everywhere. Taking this distribution into the Poisson equation (Eq. (\ref{P})), we can obtain the PB equation given in Eq. (\ref{PB}). In addition to describing the equilibrium conditions of solvated biomolecular systems, PNP equations can also model electrodiffusion processes, such as electrodiffusion reactions of mobile ions and charged ligands, ion transport across channels$^{[7, 8, 10, 17]}$.
\begin{figure}[htbp]
\begin{center}
\subfigure[]{\includegraphics[scale=0.5]{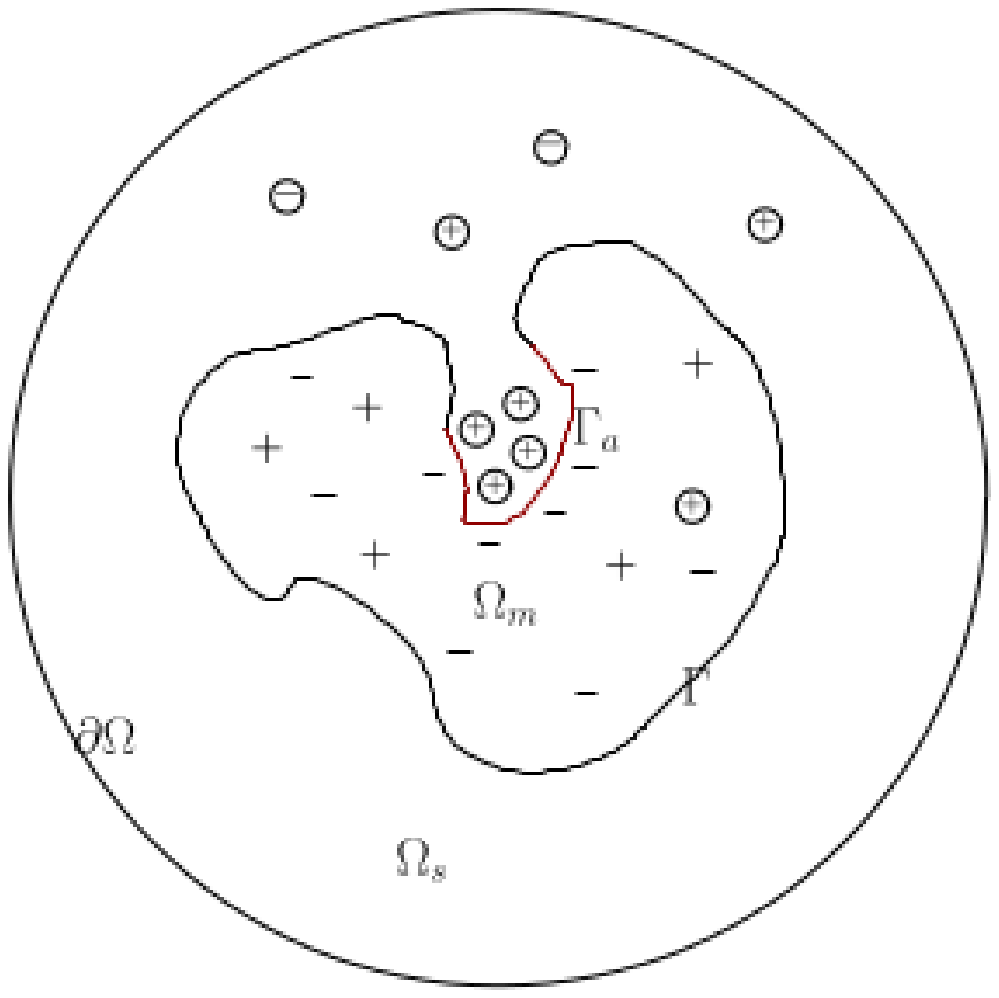}}
\subfigure[]{\includegraphics[scale=0.54]{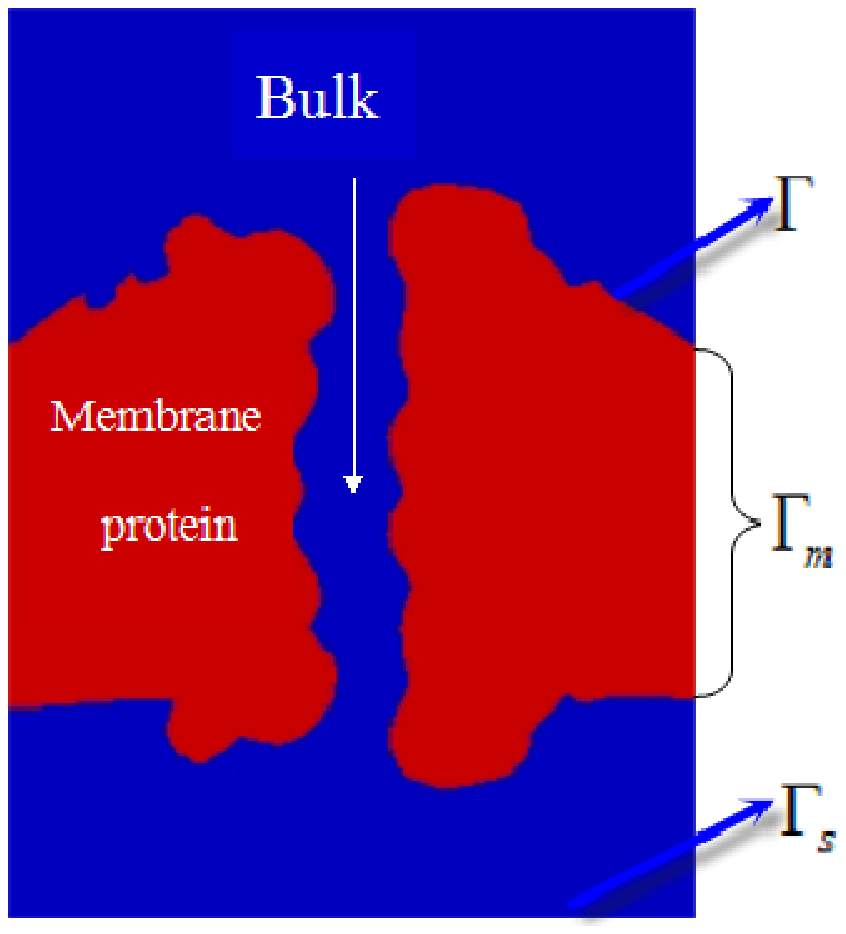}}
\end{center}
\caption{Illustrations of (a) a solvated biomolecule with active sites and (b) a channel system.}
\label{act_cha}
\end{figure}
Schematic illustrations of electro-diffusion processes are presented in Fig. \ref{act_cha}. Similar to the PB model, all the mobile ions and charged ligands are modeled as diffusive particles with vanishing sizes. A small patch $\Gamma_{a}$ of $\Gamma$ is specified to model the active site where the chemical reaction occurs and the zero Dirichlet boundary condition is set for the reactive ion while the other boundary conditons are the same as the equilibrium boundary settings$^{[6, 10]}$. The solvated system is constrained in a spherical cavity region, whereas the channel system is simulated in a cubic box. We use $\Gamma$ to denote the interface between the two regions and $\Gamma_{m}$ to denote the membrane boundary on the simulation box$^{[7]}$.

In biomolecular systems, the concrete form of electrostatic free energy is defined by
\begin{equation}
    G_{sys}=\int \Bigg\{-\frac{1}{2}\epsilon|\nabla\phi|^{2}+\rho\phi +\beta^{-1}\sum_{i=1}^{K}c_{i}\bigg[\ln (\Lambda_{i}^{3}c_{i})-1\bigg]-\sum_{i=1}^{K}c_{i}\mu_{i}\Bigg\}dV, \label{G_sys}
\end{equation}
where $\rho=\rho^{f}+\rho^{ion}$, $\Lambda_{i}$ is the thermal de Broglie wavelength, and $\mu_{i}$ is the chemical potential for the $i$th ion species. The potential in this functional is determined from the constraint of Poisson equation defined in Eq. (\ref{P}); thus, the functional depends only on ion concentrations. At equilibrium conditions that minimize the free energy, we can derive the Boltzmann distribution and the chemical potential, resulting in the PB equation$^{[32, 33]}$ and the free energy form given in Ref. 4.

\section{Improvement of continuum model}

\subsection{Size-modified PB/PNP model}  
To consider ionic size effects in ionic liquids, Andelman et al. modified the free energy form by introducing an additional solvent entropy term representing the unfavorable energy modeling the overpacking or crowding of ions and solvent molecules$^{[14]}$:
\begin{eqnarray}
    G_{sys}&=&\int\Bigg\{\frac{1}{2}\rho\phi+\beta^{-1}\sum_{i=1}^{K}c_{i}\bigg[\ln(c_{i}a_{i}^{3})-1\bigg]-\sum_{i=1}^{K}c_{i}\mu_{i}+\frac{\beta^{-1}}{a_{0}^{3}}\bigg(1-\sum_{i=1}^{K}c_{i}a_{i}^{3}\bigg) \nonumber \\
           & &\times\bigg[\ln(1-\sum_{i=1}^{K}c_{i}a_{i}^{3})-1\bigg]\Bigg\}dV, \label{G_PNP}
\end{eqnarray}
where $a_{0}$ is the size of water molecule and $a_{i}$ is the size of the $i$th ion species. In case of uniform ionic size (supposing that the sizes of different ion species and water molecules are identical), the ion concentrations at equilibrium state can be expressed as explicit functions of electrostatic potential through variation of free energy with respect to concentration, which then leads to a size-modified PB model (SMPB)$^{[14]}$. In a 1:z asymmetric solution with no fixed charge, the SMPB equation is as follows:
\begin{eqnarray}
	\epsilon\nabla^{2}\phi=-ze_{c}c_{b}\frac{e^{ze_{c}\beta\phi}-e^{-e_{c}\beta\phi}}{1-\phi_{0}+\phi_{0}(e^{ze_{c}\beta\phi}+ze^{-e_{c}\beta\phi})/(z+1)}.
\end{eqnarray}
Andelman et al. proved that the modified PB equation can be obtained from both the mean-field approximation of the partition function in a formal lattice gas formalism and the including of an entropy term for the solvent molecules in the free energy in an alternative phenomenological approach$^{[34]}$. A great decrease in counterion concentration is shown in their numerical simulations for a planar surface model and oversaturation phenomenon is prohibited. Fenley's team studied this SMPB model through comparing predictions between this equation and the nonlinear PB (NPB) equation for a low-dielectric charged spherical cavity in an aqueous salt solution and investigating the sensitivities to parameterization$^{[16, 35]}$.

To account for different ionic sizes, Chu et al. extended this model to include two different sizes and gave an explicit SMPB form in the study of ion binding to DNA duplexes$^{[15]}$. All the explicit forms are obtained by substituting the ion concentration of Boltzmann distribution in the PB equation with an explicit form of modified concentration after considering ionic volume exclusion. However, an explicit form of an ion concentration as a function of potential and ionic sizes does not exist for a general case of various ionic and solvent molecular sizes. Because the PB model (equilibrium description) can be considered as a specific case of the PNP model (non-equilibrium description), we can resort to the size-modified PNP framework to implicitly obtain the SMPB results. We have modified the steady-state PNP equations to obtain nonuniform size-modified PNP (SMPNP) equations that naturally treat arbitrary number of ion species with different sizes and can describe both equilibrium state (i.e. PB case) and non-equilibrium processes of ionic solutions$^{[10, 17]}$. The chemical potential from variation of the free energy in Eq. (\ref{G_PNP}) is used for constructing the ion flux from constitutive relations between flux and chemical potentials:
\begin{eqnarray}
	J_i=-m_ic_i\nabla\mu_i=-\frac{D_i}{k_B T}c_i\nabla\mu_i,
\end{eqnarray}
where $m_i$ is the mobility of the $i$th ion species. Substituting this $J_i$ in the NP equations given by Eq. (\ref{NP}), we obtain modified NP equations of an SMPNP model:
\begin{eqnarray}
    &\frac{\partial c_{i}(\bm{x})}{\partial t}=\nabla\cdot D_{i}(\bm{x})\bigg(\nabla c_{i}(\bm{x})+\frac{k_{i}c_{i}(\bm{x})}{1-\sum_{l}a_{l}^{3}c_{l}(\bm{x})}\sum_{l}a_{l}^{3}\nabla c_{l}(\bm{x})+\beta c_{i}(\bm{x})q_{i}\nabla\phi(\bm{x})\bigg),\text{in $\Omega_{s}$}, i=1,\ldots,K,
\end{eqnarray}
where $k_{i}=\frac{a_{i}^{3}}{a_{0}^{3}}$, and the Poisson equation remains the same as given in Eq. (\ref{P}).

\subsection{Coupling of classical density functional theory and the PNP model}
Classical DFT, a systematical theory to investigate inhomogeneous fluids, can be coupled with the PNP equations to improve simulation results for biomolecular systems. The major issue with DFT lies in the construction of excess Helmholtz free energy$^{[36]}$. The fundamental measure theory (FMT), derived by Rosenfeld, provides an excellent suggestion on how to construct the excess Helmholtz free energy for inhomogeneous hard sphere mixtures and has been developed and widely used in many studies$^{[21, 37-45]}$. According to the theory$^{[37]}$, the excess Helmholtz free energy due to hard sphere repulsion is suggested as follows:
\begin{eqnarray}
	\mathscr{F}_{ex}\Big[\big\{c_i(\bm{x})\big\}\Big]&=&\beta^{-1}\int d\bm{x} \Phi\Big[\big\{n_\alpha(\bm{x})\big\}\Big], \label{EH}\\
	\Phi\Big[\big\{n_\alpha(\bm{x})\big\}\Big]&=&-n_0\ln(1-n_3)+\frac{1}{1-n_3}(n_1n_2-\bm{n}_{V1}\cdot\bm{n}_{V2})+\frac{n_2}{24\pi(1-n_3)^2}\nonumber \label{Phi}\\
							&&\times(n_{2}^{2}-3\bm{n}_{V2}\cdot\bm{n}_{V2}), \\
        n_{\alpha}(\bm{x})&=&\sum_{i}\int c_i(\bm{x'})\omega_i^{(\alpha)}(\bm{x}-\bm{x'})d\bm{x'},
\end{eqnarray}
where $\{\omega_{i}^{(\alpha)}(\bm{x})\}$ is a set of characteristic (weight) functions for $\alpha=0,1,2,3,V1,V2$. For a three-dimensional sphere of radius $R_i$, these functions are defined by:
\begin{eqnarray}
    \omega_i^{(3)}(\bm{x})&=&\theta(|\bm{x}|-R_i),\nonumber\\
    \omega_i^{(2)}(\bm{x})&=&\delta(|\bm{x}|-R_i), \nonumber\\
    \omega_i^{(1)}(\bm{x})&=&\omega_i^{(2)}(\bm{x})/4\pi R_i, \nonumber\\
    \omega_i^{(0)}(\bm{x})&=&\omega_i^{(2)}(\bm{x})/4\pi R_i^2, \nonumber\\
    \bm{\omega}_i^{(V2)}(\bm{x})&=&\frac{\bm{x}}{|\bm{x}|}\delta(|\bm{x}|-R_i),\nonumber\\
    \bm{\omega}_i^{(V1)}(\bm{x})&=&\bm{\omega}_i^{(V2)}(\bm{x})/4\pi R_i.\nonumber
\end{eqnarray}
$\theta(x)$ is the unit step function in the definition of
\begin{equation}
  \theta(x)=
   \begin{cases}
   0 &\mbox{if $x>0$},\\
   1 &\mbox{if $x\leq0$},
   \end{cases} \nonumber
\end{equation}
and $\delta(x)$ is the Dirac delta function.

In addition to the ideal chemical potential $\mu_i^{id}=q_{i}\phi+\beta^{-1}\ln \Lambda_i^3c_i$, an excess chemical potential derived from variation of the excess free energy defined in Eqs. (\ref{EH}) and (\ref{Phi}) is introduced to account for hard sphere interactions in the form of
\begin{eqnarray}
    \mu_i^{ex}(\bm{x})=\frac{\delta\mathscr{F}_{ex}}{\delta c_i}=\beta^{-1}\int d\bm{x'}\sum_\gamma \frac{\partial\Phi}{\partial n_\gamma}\Big[\{n_\alpha(\bm{x})\}\Big] \omega_i^{(\alpha)}(\bm{x}-\bm{x'}). \label{mu_ex}
\end{eqnarray}
Ion flux is then modified to $J_{i}=-\frac{D_i}{k_B T}c_i\nabla(\mu_i^{id}+\mu_i^{ex})$ and the coupling finally yields the following modified NP equations:
\begin{eqnarray}
	&\frac{\partial c_{i}(\bm{x})}{\partial t}=\nabla\cdot D_i(\bm{x})\Big(\nabla c_i(\bm{x})+\beta q_{i}c_i(\bm{x})\nabla\phi(\bm{x})+\beta c_i(\bm{x})\nabla\mu_i^{ex}(\bm{x})\Big), \text{in $\Omega_{s}$}, i=1,\cdots,K.  \label{LHSNP}
\end{eqnarray}
It is clear that integro-differential equations are evoked when FMT is incorporated into the PNP equations, which contains a large number of freedoms and is quite difficult for numerical implementation in real biomolecular simulations. For simplicity, most numerical experiments are constrained to the 1D case or use the definition of the Dirac delta function and change of variables to transform these 3D integrals into 2D integrals on spheres to remove the singularity in the integrands$^{[20, 21, 24]}$. Only a few studies have provided 3D algorithms and performed calculations on 3D cases without biomolecules through fast Fourier transform to cope with the convolutional integrations$^{[22, 23]}$.

In contrast to the integral form of excess chemical potential and the abovementioned efforts to solve the integro-differential equations, Liu et al. have derived a local hard sphere description for $\mu_{i}^{ex}$ in the 1D case, given by
\begin{eqnarray}
	\beta\mu_{i}^{LHS}(x)=-\ln\Bigg(1-\sum_{j=1}^{K}d_{j}c_{j}(x)\Bigg)+\frac{d_{i}\sum_{j=1}^{K}c_{j}(x)}{1-\sum_{j=1}^{K}d_{j}c_{j}(x)},
\end{eqnarray}
where $d_{j}$ is the diameter of the $j$th ion species$^{[45]}$.
To obtain more accurate simulations of biomolecular systems, we have incorporated a local excess chemical potential originating from 3D hard sphere interactions$^{[27]}$
\begin{eqnarray}
	\beta\mu_i^{LHS}(\bm{x})&=&-\ln\big(1-\sum_j\frac{4}{3}\pi R_j^3 c_j(\bm{x})\big)
			+\frac{R_i\sum_j 4\pi R_j^2 c_j(\bm{x})}{1-\sum_j\frac{4}{3}\pi R_j^3 c_j(\bm{x})} \nonumber \\
			 & &  +\frac{4\pi R_i^2\sum_j R_j c_j(\bm{x})}{1-\sum_j\frac{4}{3}\pi R_j^3 c_j(\bm{x})}
		  +\frac{4}{3}\pi\frac{R_i^3\sum_j c_j(\bm{x})}{1-\sum_j\frac{4}{3}\pi R_j^3 c_j(\bm{x})}\label{muLHS}
\end{eqnarray}
into the PNP model to form a local hard sphere PNP (LHSPNP) model that can capture the ionic finite size and excluded volume effects featured by the saturation phenomenon but simplify numerical calculations at the same time. The first term in Eq. (\ref{muLHS}) is similar to the excess chemical potential in the SMPNP model where $\beta\mu_{i}^{ex}(\bm{x})=-k_{i}\ln\big(1-\sum_j a_j^3 c_j(\bm{x})\big)$. Furthermore, similarities between the numerical results of these two models are also observed under most boundary conditions.

In addition to chemical potential due to hard sphere repulsions, chemical potential terms resulting from electrostatic correlations and short-range attraction interactions has also been analyzed in many studies$^{[22, 23, 42]}$. There have also been some improvements in the FMT to account for discrete particle interactions. Kierlik and Rosinberg proposed a simplified version of the FMT, which was proven to be equivalent to the original FMT but required only four scalar weight functions$^{[44]}$. Using the Mansoori-Carnahan-Starling-Leland bulk equation of state, Roth et al. and Wu et al. have independently, published the white-bear version or the modified FMT to make simulation results more accurate$^{[39, 41]}$. For inhomogeneous fluids of nonspherical hard particles, Mecke et al. have derived a fundamental measure theory using the Gauss-Bonnet theorem$^{[43]}$. These different forms of excess Helmholtz free energy lead to different expressions of excess chemical potential and thus different coupling results of DFT and PNP.

\subsection{Coupling polar and nonpolar interactions}
Polar and nonpolar interactions are coupled. In recent years, a few groups used a variational framework to study the coupled system$^{[28-30, 46]}$. In particular, in the studies pertaining to Refs. 28, 29 and 30 the biomolecular surface (interface between solute and solvent) was not fixed, but was determined as a model output. A variational model is proposed under the consideration of an assembly of solutes with arbitrary shape and composition surrounded by a dielectric solvent in a macroscopic volume to explicitly couple hydrophobic, dispersion, and electrostatic contributions in a continuum model. Typical free energy, including volume energy, surface energy, van der Waals interaction energy and continuum electrostatic free energy, is expressed as a functional of a characteristic function $\nu$
\begin{eqnarray}
  G[\nu]&=&\int_{\Omega}\Bigg\{P\Big(1-\nu(\bm{x})\Big)+\gamma(\bm{x})|\nabla\nu(\bm{x})|+U_{VDW}(\bm{x}, c_w(\bm{x}), \{c_i(\bm{x})\})+\frac{\epsilon(\bm{x})}{2}|\nabla\phi(\bm{x})|^2 \nonumber \\
  &&+\sum_{i=1}^{K}c_{i}(\bm{x})\Big(\ln\big(\Lambda_{i}^{3}c_{i}(\bm{x})\big)-1\Big)\Bigg\}d\bm{x}.
\end{eqnarray}
$P$ is the pressure difference between the liquid and vapor, $\nu(\bm{x})$ is 0 in space empty of solvent and is 1 elsewhere, $U_{VDW}(\bm{x}, c_w(\bm{x}), \{c_i(\bm{x})\})$ is the solvent-solute and/or solvent-solvent van der Waals interactions (model dependent), and $c_w(\bm{x})$ is the water molecular concentration. Optimal volume and surface are obtained via minimization of the free energy. Li et al. provided a representation of the interface by a phase field and expressed the free energy by all possible phase fields$^{[29]}$. The equilibrium conformations and free energies of an underlying molecular system are determined by the variational principle. Wei used the differential geometry theory of surfaces for geometric separation of the macroscopic domain of the solvent from the microscopic domain of the solute, and for dynamical coupling of the continuum and discrete descriptions, to investigate multiscale multiphysics and multidomain models$^{[30]}$. The free energy functional can also properly include other energy terms from quantum mechanics, fluid mechanics, molecular mechanics, and elastomechanics, which will lead to more complicated models.

\subsection{Other improvements}
In addition to the above improvements, some other modifications have also been made to the continuum model. The homogeneous assumption of dielectric medium does not coincide with experimental results of salt water solutions$^{[47]}$. We explored a variable dielectric PB (VDPB) model by considering the dielectric as an explicit function of ionic sizes and concentrations$^{[25]}$. Andelman et al. proposed a dipolar PB (DPB) model by coarse graining the interaction of individual ions and dipoles interacting together$^{[48]}$. A Poisson-Fermi model is derived by introducing a Landau-Ginzburg-like free energy functional to describe the interplay between overscreening and crowding$^{[13, 19]}$. Other modifications, for example, incorporating image charge effects$^{[49]}$ and Lennard-Jones hard sphere repulsion$^{[24]}$ into the primitive model, have also been studied to improve the continuum model. Horng et al. introduced the finite size effect by treating ions and side chains as solid spheres and using Lennard-Jones hard sphere repulsion potentials to characterize this effect, and observed ion-selectivity behavior in a simple 1D analysis and simulation$^{[46]}$.

Other beyond-mean-field models have been developed at the same time to account for ion-ion correlations. Kjellander et al. introduced the dressed ion theory--a formally exact theory for electrolyte solutions and colloid dispersions by writing the ion-ion correlation function as a sum of a short-range and a (relatively) long-range part$^{[50, 51]}$. For nucleic acid mixtures, the tightly bound ion theory was presented by Tan et al., which captures the fluctuations and the electrostatic and ion-ion correlations and provides reliable predictions for ion distribution and thermodynamic properties$^{[52, 53]}$. The integral equation theory is used to evaluate distribution functions of solvents and calculate the liquid structure and thermodynamic properties$^{[54-57]}$. In case of high valence ions, a strong coupling theory as a correction to the PB approach is available to predict counterion distributions around charged objects$^{[58]}$. Furthermore, a system of self-consistent partial differential equations is derived to extend the PB equation to include electrostatic correlation and fluctuation effects$^{[12, 59]}$ and recently a numerical method was developed to solve the equations$^{[60]}$.

\section{Numerical progress}
In the simulation of biomolecular systems, progress has been made for developing more accurate and efficient algorithms$^{[1, 31]}$. In the finite difference method for solution of the PB equation, to enforce the interface conditions and to obtain better numerical convergence, Wei et al. implemented the matched interface and boundary (MIB) method to accurately treat the interface conditions, and also used the regularization scheme described in Ref. 61 to remove the charge singularities in the molecule to achieve a higher level of accuracy$^{[62, 63]}$. Luo's group developed the immersed interface method (IIM), a new discretization method, and implemented it in the PBSA solver that is incorporated in the Amber package$^{[64]}$. Lu et al. used the finite element method and a body-fitted molecular surface mesh to solve the interface problem with a high level of accuracy$^{[1, 7]}$. At the same time, considerable efforts have been made to increase the speed of computations. Xie et al. employed the MPI to present a parallel adaptive finite element algorithm to solve the 3D electro-diffusion equations$^{[8]}$. Geng et al. provided a GPU-accelerated direct-sum boundary integral method to solve the linear PB equation$^{[65]}$. Yokota et al. presented a GPU-accelerated algorithm with the fast multipole method in conjunction with a boundary element method (BEM) formulation, resulting in strong scaling with parallel efficiency of 0.5 for 512 GPUs$^{[66]}$. Lu et al. parallelized the adaptive fast multipole PB (AFMPB) package with the Cilk Plus system to harness the computing power of both multicore and vector processing, which achieves nearly optimal computational complexity in BEM and successfully solves the PB equation on a workstation for macromolecular systems such as a virus with tens of million degrees of freedom$^{[67]}$.

\section{Conclusions}
We have summarized the PB/PNP types of continuum models based on mean field approximation, their improvements to include ionic size effects, other physical properties that have been missed in previous models, the coupling of polar and nonpolar interactions, and numerical progress for biomolecular simulations. The main improvements for including size effects were achieved by introducing an excess chemical potential term to account for discrete particle interactions ignored in the continuum model. However, differing from the SMPB/SMPNP models which are still mean field approximations, coupling of DFT and PNP provides a reliable framework for particle interactions. Variational treatment of the biomolecular surface is included in the free energy that couples polar and nonpolar interactions to obtain both an optimal surface and a minimum free energy. Numerical progress makes it possible to obtain more accurate and efficient algorithms for simulating biomolecular systems. It is expected that much efforts will be made in the future for developing beyond-mean-field continuum models that can not only capture important detailed physical effects but also keep numerically tractable for biomolecular systems. At the same time, effective numerical techniques (along with advances in modern computer science and mathematics) need to be developed for dealing with larger and more complicated simulation systems.

\end{document}